\documentclass[prb,twocolumn,superscriptaddress,showpacs]{revtex4}

\usepackage[dvips]{graphicx}
\usepackage{amsmath}

\newcommand{\Tr}{\mathop{\mathrm{Tr}}} 
\newcommand{\iu}{i}                    
\newcommand{\upd}{d}                   
\newcommand{\mat}{\boldsymbol}         
\newcommand{\im}{\textrm{Im}}          

\newcommand{\gc}{G}
\newcommand{\gf}{\mathcal{G}}
\newcommand{\gb}{g}
\newcommand{\igc}{F}
\newcommand{\igf}{\mathcal{F}}

\newcommand{\se}{\Sigma}
\newcommand{\ga}{\Gamma}

\newcommand{\ttt}{\gamma}
\newcommand{\ttu}{\eta}

\newcommand{\Ep}{E_+}
\newcommand{\td}{\tilde}

\begin{document}

\title{Modeling elastic and photoassisted transport in organic
molecular wires: length dependence and current-voltage characteristics}

\author{J. K. Viljas}
\email{janne.viljas@kit.edu}
\affiliation{Institut f\"ur Theoretische Festk\"orperphysik
and DFG-Center for Functional Nanostructures,
Universit\"at Karlsruhe, D-76128 Karlsruhe, Germany}
\affiliation{Forschungszentrum Karlsruhe,
Institut f\"ur Nanotechnologie, D-76021 Karlsruhe, Germany }

\author{F. Pauly}
\affiliation{Institut f\"ur Theoretische Festk\"orperphysik
and DFG-Center for Functional Nanostructures,
Universit\"at Karlsruhe, D-76128 Karlsruhe, Germany}
\affiliation{Forschungszentrum Karlsruhe,
Institut f\"ur Nanotechnologie, D-76021 Karlsruhe, Germany }

\author{J. C. Cuevas}
\affiliation{Departamento de F\'{\i}sica Te\'orica de la Materia
Condensada, Universidad Aut\'onoma de Madrid, E-28049 Madrid, Spain}
\affiliation{Institut f\"ur Theoretische Festk\"orperphysik
and DFG-Center for Functional Nanostructures,
Universit\"at Karlsruhe, D-76128 Karlsruhe, Germany}
\affiliation{Forschungszentrum Karlsruhe,
Institut f\"ur Nanotechnologie, D-76021 Karlsruhe, Germany }

\date{\today}

\begin{abstract}
  
  Using a $\pi$-orbital tight-binding model, we study the elastic and
  photoassisted transport properties of metal-molecule-metal junctions
  based on oligophenylenes of varying lengths. The effect of
  monochromatic light is modeled with an ac voltage over the
  contact. We first show how the low-bias transmission function can be
  obtained analytically, using methods previously employed for simpler
  chain models. In particular, the decay coefficient of the
  off-resonant transmission is extracted by considering both a
  finite-length chain and infinitely extended polyphenylene. Based on
  these analytical results, we discuss the length dependence of the
  linear-response conductance, the thermopower, and the light-induced
  enhancement of the conductance in the limit of weak intensity and
  low frequency. In general the conductance-enhancement is calculated
  numerically as a function of the light frequency. Finally, we
  compute the current-voltage characteristics at finite dc voltages,
  and show that in the low-voltage regime, the effect of low-frequency
  light is to induce current steps with a voltage separation
  determined by twice the frequency. These effects are more pronounced
  for longer molecules.  We study two different profiles for the dc
  and ac voltages, and it is found that the results are robust with
  respect to such variations.  Although we concentrate here on the
  specific model of oligophenylenes, the results should be
  qualitatively similar for many other organic molecules with a large
  enough electronic gap.
 
\end{abstract}



\maketitle


\section{Introduction} \label{s.intro}

The use of single-molecule electrical contacts for optoelectronic
purposes such as light sources, light sensors, and photovoltaic
devices is an exciting idea. Yet, due to the difficulties that
light-matter interactions in nanoscale systems pose for theoretical
and experimental investigations, the possibilities remain largely
unexplored.  Concerning experiments, it has been shown that light can
be used to change the conformation of some molecules even when they
are contacted to metallic electrodes, thus enabling light-controlled
switching.\cite{vanderMolen06} Some evidence of photoassisted
processes influencing the conductance of laser-irradiated metallic
atomic contacts has also been obtained.\cite{Guhr07a} Theoretical
investigations of light-related effects in molecular contacts are more
numerous,
\cite{Buker02,Tikhonov02a,Tikhonov02b,Lehmann02b,Urdaneta03,Kohler04,Urdaneta05,Galperin05,Galperin06,Bittner05,Welack06,Harbola06,Viljas07a,Viljas07b,Liu07,Li07,Orellana07}
but they are mostly based on highly simplified models, whose validity
remains to be checked by more detailed
calculations\cite{Kurth05,Galperin07} and experiments.  However, for
the description of the basic phenomenology, model approaches can be
very fruitful, as they have been in studies of elastic transport in
the past. Properties of linear single-orbital tight-binding (TB)
chains, in particular, have been studied in detail, and to a large
part analytically.
\cite{McConnell61,Mujica94a,Mujica94b,Mujica96,Nitzan01b,Segal03,Kohler04,Asai05,Segal05,Painelli06,Maiti07,Tomfohr02}
In a step towards a more realistic description of the geometry,
symmetries, and the electronic structure of particular molecules,
empirical TB approaches such as the (extended) H\"uckel method have
proved
useful.\cite{Tikhonov02b,Samanta96,Dalgleish06,Viljas07a,Stafford07}

Based on a combination of density-functional calculations and simple
phenomenological considerations, we have recently described the
photoconductance of metal-oligophenylene-metal
junctions.\cite{Viljas07b} It was discussed how the linear-response
conductance may increase by orders of magnitude in the presence of
light.  This effect can be seen as the result of a change in the
character of the transport from off-resonant to resonant, due to the
presence of photoassisted
processes.\cite{Tikhonov02a,Tikhonov02b,Viljas07b} Consequently, the
decay of the conductance with molecular length is slowed down,
possibly even making the conductance
length-independent.\cite{Tikhonov02b,Viljas07b}

In this paper we apply a H\"uckel-type TB model of
oligophenylene-based contacts\cite{Pauly07b}
combined with Green-function methods\cite{Viljas07a} to study the
effects of monochromatic light on the dc current in
metal-oligophenylene-metal contacts.  Again we concentrate on the
dependence of these effects on the length of the molecule.  We begin
with a detailed account of the elastic transport properties of the
model, and show that the zero-bias transmission function can be
obtained analytically, similarly to simpler chain
models.\cite{Mujica94a,Asai05} We demonstrate how information about
the length dependence of the transmission function for a finite wire
can be extracted from an infinitely extended polymer.  Based on these
analytical results, we discuss the length dependences of the
conductance and the photoconductance for low-intensity and
low-frequency light. While the conductance decays
exponentially with length, its relative enhancement due to light exhibits a
quadratic behavior.  Here we also briefly consider the thermopower,
whose length dependence is linear.  Next, we calculate numerically the
zero-bias photoconductance as a function of the light frequency
$\omega$, and find that the conductance-enhancement due to light is
typically very large.\cite{Tikhonov02b,Kohler04,Viljas07b} In
particular, we show that the results of Ref.\ \onlinecite{Viljas07b}
are expected to be robust with respect to variations in the assumed
voltage profiles. Finally, we describe how the step-like
current-voltage ($I$-$V$) characteristics are modified by light. 
At high $\omega$ the most obvious effect is the
overall increase in the low-bias current. At low $\omega$, additional
current steps similar to those in microwave-irradiated superconducting
tunnel junctions\cite{Dayem62,Tien63} can be seen. Their separation, in
our case of symmetric junctions, is roughly $2\hbar\omega/e$.

TB models of the type we shall consider neglect various interaction
effects (see Sec.\ \ref{s.discussion} for a discussion), and thus
cannot be expected to give quantitative predictions. However, the
qualitative features of the results rely only on the tunneling-barrier
character of the molecular contacts, which results from the fact that
the Fermi energy of the metal lies in the gap between the
highest-occupied and lowest-unoccupied molecular orbitals (HOMO and
LUMO) of the molecule.  Thus, these features should remain similar for
junctions based on many other organic molecules exhibiting large
HOMO-LUMO gaps. The light-induced effects, if verified experimentally,
could be used for detecting light, or as an optical gate (or ``third
terminal'') for purposes of switching.

The rest of the paper is organized as follows. In Sec.\ \ref{s.gene}
we describe our theoretical approach, discuss the general properties
of TB wire models, and introduce the Green-function method for the
calculation of the elastic transmission function. Then, in Sec.\
\ref{s.phenyl} we calculate the transmission function of
oligophenylene wires analytically. The decay coefficient for the
off-resonant transmission is extracted also from infinitely extended
polyphenylene. Following that, in Sec.\ \ref{s.results} we present our
numerical results for the conductance, the thermopower, the
photoconductance, and the $I$-$V$ characteristics.  Finally, Sec.\
\ref{s.discussion} ends with our conclusions and some discussion.
Details on the calculation of the time-averaged current in the
presence of light are deferred to the appendixes.  In App.\
\ref{s.simplecurr} a simplified interpretation of the current formula
is derived, and in App.\ \ref{s.lightcurr} a brief account of the
general method is given. Readers mainly interested in the discussion
of the results for the physical observables can skip most of Secs.\
\ref{s.gene} and \ref{s.phenyl}, and proceed to Sec.\ \ref{s.results}.


\section{Theoretical framework} \label{s.gene}

\subsection{Transport formalism}

Our treatment of the transport characteristics for the two-terminal
molecular wires is based on Green's functions and the
Landauer-B\"uttiker formalism, or its generalizations.  Assuming the
transport to be fully elastic, the dc electrical current through a
molecular wire can be described with
\begin{equation} \label{e.curr}
I(V)=\frac{2e}{h}\int\upd E\tau(E,V)
[f_L(E)-f_R(E)].
\end{equation}
Here $V$ is the dc voltage and $\tau(E,V)$ is the voltage-dependent
transmission function, while $f_X(E)=1/[\exp((E-\mu_X)/k_BT_X)+1]$,
$\mu_X$, and $T_X$ are the Fermi function, the electrochemical
potential, and the temperature of side $X=L,R$,
respectively.\cite{Note2} The electrochemical potentials satisfy
$eV=\Delta\mu=\mu_L-\mu_R$, and we can choose them symmetrically as
$\mu_{L}=E_F+eV/2$ and $\mu_{R}=E_F-eV/2$, where $E_F$ is the Fermi
energy. For studies of dc current we always assume $T_L=T_R=0$.  Of
particular experimental interest is the linear-response conductance
$G_{dc}=\partial I/\partial V|_{V=0}$, given by the Landauer formula
$G_{dc}=G_0\tau(E_F)$, where $G_0=2e^2/h$ and $\tau(E)=\tau(E,V=0)$.
In most junctions based on organic oligomers, the transport can be
described as off-resonant tunneling. This results in the well-known
exponential decay of $G_{dc}$ with the number $N$ of monomeric
units in the molecule.\cite{Akkerman08} At finite voltages $V$, the
current increases in a stepwise manner as molecular levels begin to 
enter the bias window between $\mu_L$ and $\mu_R$ (Ref.\ 
\onlinecite{Mujica96}). We shall consider both of these phenomena below.

If a small temperature difference $\Delta T=T_L-T_R$ at an average
temperature $T=(T_L+T_R)/2$ is applied, heat currents and
thermoelectric effects can arise.\cite{Paulsson03,Reddy07,Pauly07b} In
an open-circuit situation, where the net current $I$ must vanish, a
thermoelectric voltage $\Delta\mu/e$ is generated to balance the
thermal diffusion of charge carriers.  In the linear-response regime
the proportionality constant $S=-(\Delta\mu/e\Delta T)_{I=0}$ is the
Seebeck coefficient.  We will briefly consider this quantity below as
an example of an observable with a linear dependence on the molecular
length $N$, but will not enter a more detailed discussion of
thermoelectricity or heat transport.

The quantity we are most interested in is the dc current in the
presence of monochromatic electromagnetic radiation, which we refer to
as light independently of its source or frequency $\omega$. We model the light
as an ac voltage with harmonic time-dependence $V(t)=V_{ac}\cos(\omega
t)$ over the contact. The current averaged over one period of $V(t)$
can be written in the form \cite{Jauho94,Kohler04,Viljas07a}
\begin{equation} \label{e.photocurr}
\begin{split}
I(V;\alpha,\omega) = &
\frac{2e}{h}
\sum_{k=-\infty}^\infty \int \upd E [
\tau_{RL}^{(k)}(E,V;\alpha,\omega) \\
& \times f_L(E)
-\tau_{LR}^{(k)}(E,V;\alpha,\omega)f_R(E)].
\end{split}
\end{equation}
Here the transmission coefficient $\tau^{(k)}_{RL}(E)$, for example,
describes photoassisted processes taking an electron from left ($L$)
to right ($R$), under the absorption of a total of $k$ photons with
energy $\hbar\omega$.  The parameter $\alpha=eV_{ac}/\hbar\omega$
describes the strength of the ac drive.\cite{Wagner97} It is
determined by the intensity of the incident light and possible
field-enhancement effects taking place in the metallic
nanocontact.\cite{Grafstrom02} Again, in addition to the full $I$-$V$
characteristics, we study in more detail the case of linear response
with respect to the dc bias, i.e., the photoconductance
$G_{dc}(\alpha,\omega)=\partial I(V;\alpha,\omega)/\partial V|_{V=0}$.
The arguments $\alpha$ and $\omega$ distinguish it from the
conductance $G_{dc}$, although we sometimes omit $\alpha$ for notational
simplicity.  The calculation of the coefficients
$\tau^{(k)}_{RL/LR}(E)$ is rather complicated in
general,\cite{Viljas07a} and we defer comments on this procedure to
App.\ \ref{s.lightcurr}. Below we shall mostly refer to an approximate
formula (see App.\ \ref{s.simplecurr}) that can be expressed in terms
of $\tau(E)$. This amounts to a treatment of the problem on the level
of the Tien-Gordon approach.\cite{Tien63,Kohler04,Platero04} The full
Green-function formalism for systems involving ac driving is presented
in Ref.\
\onlinecite{Viljas07a}.

In noninteracting (non-self-consistent) models it is in general not
clear how the voltage drop should be divided between the different
regions of the wire, and the electrode-wire interfaces. A
self-consistent treatment would be in order in particular for
asymmetrically coupled molecules. We only concentrate on left-right
symmetric junctions, where where both the dc and ac voltages ($V$ and
$V_{ac}$) are assumed to drop according to one of two different
symmetrical profiles. The symmetry of the junctions excludes
rectification effects, such as light-induced dc
photocurrents in the absence of a dc bias
voltage.\cite{Grafstrom02,Kohler04,Galperin05} However, light can still have a
strong influence on the transmission properties of the molecular
contact, as will be discussed below.  It will be shown that
our conclusions are essentially independent of the assumed voltage
profile.

\subsection{Wire models}

\begin{figure}[!tb]
\includegraphics[width=0.85\linewidth,clip=]{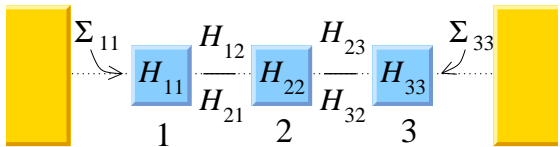}
\caption{(Color online) A finite block chain of length $N$=3 connected
to electrodes at its two ends. This gives rise to self energies
$\mat{\se}_{11}$ and $\mat{\se}_{NN}$ on the terminating blocks.  }
\label{f.blockjunction}
\end{figure}
Below we will specialize to the case of a metal-oligophenylene-metal
junction.  However, to make some general remarks, let us first
consider a larger class of molecular wires that can be described as
$N$ separate units forming a chain, where only the nearest neighbors
are coupled (see Fig.\ \ref{f.blockjunction}). We only discuss the
calculation of the elastic transmission function $\tau(E,V)$ here, as
this will be the focus of our analytical considerations in Sec.\
\ref{s.phenyl}. From this quantity (at $V=0$), the various
linear-response coefficients such as the conductance and the
thermopower can be extracted.  Furthermore, as already mentioned, it
suffices for an approximate treatment of the amplitudes
$\tau_{RL}^{(k)}(E)$ as well.

We assume a basis $|\chi^{(\alpha)}_p\rangle$ of local (atomic)
orbitals, where $p=1,\ldots,N$ indexes the unit, while
$\alpha=1,\ldots,M_p$ denotes the orbitals in each
unit.\cite{Note5}  For simplicity, the basis is taken to be orthonormal, i.e.
$\langle\chi^{(\alpha)}_p|\chi^{(\beta)}_q\rangle=\delta_{\alpha\beta}\delta_{pq}$.
The (time-independent) Hamiltonian
$H^{(\alpha,\beta)}_{pq}=\langle\chi^{(\alpha)}_p|\hat{H}|\chi^{(\beta)}_q\rangle$
of the wire is then of the block-tridiagonal form
\begin{equation} \label{e.fullham}
\begin{split}
\mat{H}= 
&\left(\begin{matrix}
\mat{H}_{11}  & \mat{H}_{12}  &        &        &        \\
\mat{H}_{21} & \mat{H}_{22}  & \mat{H}_{23}  &        &        \\
       & \ddots & \ddots & \ddots &        \\
       &        & \mat{H}_{N-1,N-2} & \mat{H}_{N-1,N-1}  & \mat{H}_{N-1,N}  \\
       &        &        & \mat{H}_{N,N-1} & \mat{H}_{NN}  \\
\end{matrix}\right),
\end{split}
\end{equation}
where $\mat{H}_{pq}$ with $p,q=1,\ldots,N$ are $M_p\times M_q$ matrices.
(The unindicated matrix elements are all zeros.)

In the non-equilibrium Green-function picture, the effect of coupling
the chain to the electrodes is described in terms of ``lead self
energies''.\cite{Dattabook} We assume these to be located only on the
terminal blocks of the chain, with components $\mat{\se}_{11}$ and
$\mat{\se}_{NN}$.  The inverse of the stationary-state retarded
propagator for the coupled chain will then be of the form
\begin{equation} \label{e.tridiag}
\mat{\igc}=
\left(\begin{matrix} 
\mat{F}_{11} & \mat{h}_{12}  &                &     &   \\
\mat{h}_{21} & \mat{h}_{22}  & \mat{h}_{23}  &              &  \\
       & \ddots & \ddots & \ddots        & \\
       &        & \mat{h}_{N-1,N-2} & \mat{h}_{N-1,N-1}  & \mat{h}_{N-1,N} \\
       &        &        & \mat{h}_{N,N-1} & \mat{F}_{NN}\\
\end{matrix}\right).
\end{equation}
Here $\mat{h}_{p,p\pm 1}=-\mat{H}_{p,p\pm 1}$,
$\mat{h}_{pp}=\Ep\mat{1}_{pp}-\mat{H}_{pp}$, and $\Ep=E+\iu 0^+$,
while $\mat{F}_{11}=\mat{h}_{11}-\mat{\se}_{11}$ and
$\mat{F}_{NN}=\mat{h}_{NN}-\mat{\se}_{NN}$.  Charge-transfer
effects between the molecule and the metallic electrodes shift the
molecular levels with respect to the Fermi energy $E_F$. In a TB
model, these can be represented by shifting the diagonal elements of
$\mat{H}$.  Once a transport voltage $V$ is applied, further shifts
are induced.  In our model the voltage-induced shifts will be taken
from simple model profiles, and the relative position of $E_F$ will be
treated as a free parameter.

Effective numerical ways of calculating the propagator
$\mat{\gc}=\mat{\igc}^{-1}$ for block-tridiagonal Hamiltonians
exist.\cite{Markussen06,Todorov96} In Sec.\ \ref{s.phenyl} we shall be
interested in a special case, where $\mat{H}_{p,p-1}=\mat{H}_{-1}$,
$\mat{H}_{p,p+1}=\mat{H}_{1}$ and $\mat{H}_{pp}=\mat{H}_{0}$ with the
same $\mat{H}_{1}=\mat{H}_{-1}^{T}$ and $\mat{H}_{0}$ (of dimension
$M_p=M$) for all $p$, describing an oligomer of identical monomeric
units. In such cases also analytical progress in calculating the
current in Eq.\ (\ref{e.curr}) may be possible.  Once the Green
function $\mat{\gc}$ is known, the transmission function is given
by\cite{Dattabook}
\begin{equation} \label{e.genetrans}
\tau(E,V)=\Tr[\mat{\ga}_{11}\mat{\gc}_{1N}\mat{\ga}_{NN}(\mat{\gc}_{1N})^{\dagger}],
\end{equation}
where $\mat{\ga}_{11}=-2\im{\mat{\se}_{11}}$ and
$\mat{\se}_{11}(E,V)=\mat{\se}_{11}(E-eV/2)$, for example.  

Typically $E_F$ lies within the HOMO-LUMO gap, resulting in the
exponential decay $\tau(E_F)\sim e^{-\beta(E_F)N}$ with $N$,
characteristic of off-resonant transport. The decay coefficient
$\beta(E_F)$ is actually independent of $\mat{\se}_{11}$ and
$\mat{\se}_{NN}$.
This can be seen by considering the Dyson equation $\mat{\gc} =
\mat{\gf}+\mat{\gf}\mat{\se}\mat{\gc}$, where $\mat{\gc}$ and
$\mat{\gf}$ are the Green function of the coupled and uncoupled wires,
respectively, and $\mat{\se}$ is the matrix for the lead
self-energies.  Assuming that $\mat{\gf}_{1N}$ decays exponentially
with $N$, then
\begin{equation}
\mat{\gc}_{1N}\approx(\mat{1}-\mat{\gf}_{11}\mat{\se}_{11})^{-1} \mat{\gf}_{1N}
\end{equation}
when $N\rightarrow\infty$, and therefore $\mat{\gc}_{1N}$ decays with
the same exponent.  Thus, one can in principle obtain the decay
exponent from the propagator of an isolated molecule, or even an
infinitely extended polymer. In the next Section we demonstrate this
by extracting the decay exponent of a finite oligophenylene junction
from the propagator for polyphenylene. We note that in doing so, we
neglect the practical difficulty of determining the correct relative
position of $E_F$.

There are efficient numerical methods for computing the lead
self-energies for different types of electrodes and various bonding
situations between them and the wire. Typically, the methods are based
on the calculation of surface Green's functions.\cite{Viljas05} Below
we shall simply treat the self-energies as parameters.


\section{Phenyl-ring-based wires} \label{s.phenyl}

In this Section we discuss a special case of the type of wire model
introduced above, describing an oligomer of phenyl rings coupled to
each other via the para ($p$) position.\cite{Pauly07b} The bias
voltage $V$ is assumed to be zero.  In the special case that we will
consider, the inversion of Eq.\ (\ref{e.tridiag}) can then be done
analytically with the subdeterminant method familiar from elementary
linear algebra.\cite{Mujica94a,Mujica94b,Mujica96,Asai05} Below, we
first use this method for calculating the propagator of the
finite-wire junction and derive the decay exponent $\beta(E)$ of the
transmission function at off-resonant energies.  After that we
rederive the decay exponent by considering an infinitely extended
polymer of phenyl rings.


\subsection{Oligo-$p$-phenylene junction} \label{s.oligophenyl}

\begin{figure}[!tb]
\includegraphics[width=0.85\linewidth,clip=]{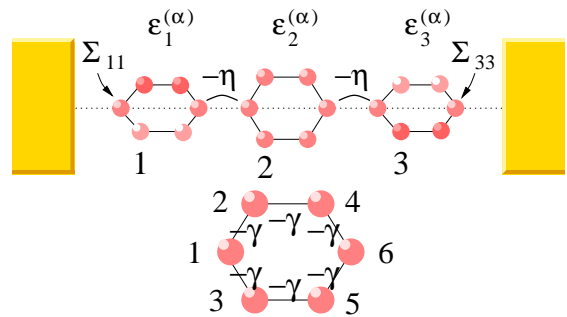}
\caption{(Color online) A finite chain of length $N$=3 connected to electrodes
at its two ends. This gives rise to self energies $\mat{\se}_{11}$ and
$\mat{\se}_{NN}$ on the end sites.  The nearest-neighbor hoppings
inside the ring ($-\ttt$) and between the rings ($-\ttu $) are
different. The lower part indicates also the numbering of the $M=6$ 
carbon atoms within a ring.  }
\label{f.phenyljunction}
\end{figure}

Our model for the oligophenylene-based molecular junction is depicted in
Fig.\ \ref{f.phenyljunction}.  Within a simple $\pi$-electron picture,
the electronic structure of the oligophenylene molecule can be
described with a nearest-neighbor TB model with two
different hopping elements $-\ttt$ and $-\ttu$ (Ref.\
\onlinecite{Note1}).  Here $-\ttt$ is for hopping within a phenyl ring,
between the $p$ orbitals oriented perpendicular to the ring plane,
while $-\ttu$ describes hopping between adjacent rings. Due to the
symmetry of the orbitals, the magnitude of $\ttu$ depends on the angle
$\varphi$ between the rings proportionally to $\cos\varphi$ (Ref.\
\onlinecite{Pauly07a}).  We shall assume that $\ttu =
\ttt\cos\varphi$, and thus $|\ttu|\leq \ttt$.  In this way the natural
energy scale of the model is set by $\ttt$ alone.

The ring-tilt angle $\varphi$ can be controlled to some extent using
side groups. For example, two side groups bonded to adjacent phenyl
rings can repel each other sterically, thus increasing the
corresponding tilt angle.\cite{Venkataraman06,Pauly07a} In fact, even
the pure oligophenylenes in the uncharged state have
$\varphi=30^\circ-40^\circ$ due to the repulsion of the hydrogen
atoms.\cite{Pauly07a,Pauly07b} However, side groups can introduce also
``charging'' or ``doping'' effects, which shift the molecular 
levels.\cite{Venkataraman07}

For definiteness, we number the $M=6$ carbon atoms of a phenyl ring 
according to the lower part of Fig.\ \ref{f.phenyljunction}.
The corresponding orbitals appear in the basis in this order.
Thus the blocks in Eq.\ (\ref{e.fullham}) are
\begin{equation} \label{e.diagblock}
\mat{H}_{q,q}=
\left(\begin{matrix}
\epsilon_{q}^{(1)} & -\ttt & -\ttt & 0 & 0 & 0 \\
-\ttt & \epsilon_{q}^{(2)} & 0 & -\ttt & 0 & 0 \\
-\ttt & 0 & \epsilon_{q}^{(3)} & 0 & -\ttt & 0 \\
0 & -\ttt & 0 & \epsilon_{q}^{(4)} & 0 & -\ttt \\
0 & 0 & -\ttt & 0 & \epsilon_{q}^{(5)} & -\ttt \\
0 & 0 & 0 & -\ttt & -\ttt & \epsilon_{q}^{(6)} \\
\end{matrix}\right)
\end{equation}
for $q=1,\ldots,N$ and 
\begin{equation} \label{e.offdiagblock}
\mat{H}_{q,q-1}=
\left(\begin{matrix}
0 & 0 & 0 & 0 & 0 & -\ttu \\
0 & 0 & 0 & 0 & 0 & 0 \\
0 & 0 & 0 & 0 & 0 & 0 \\
0 & 0 & 0 & 0 & 0 & 0 \\
0 & 0 & 0 & 0 & 0 & 0 \\
0 & 0 & 0 & 0 & 0 & 0 \\
\end{matrix}\right),
\end{equation}
with $\mat{H}_{q+1,q}=[\mat{H}_{q,q+1}]^T$.  Here the onsite energies
$\epsilon_q^{(\alpha)}$ may be shifted non-uniformly to describe effects
of possible side-groups.\cite{Pauly07b} For simplicity, we shall
consider all phenyl rings to have a similar chemical environment, and
thus all onsite energies are taken to be equal.

As a first step we note that, assuming
$\epsilon_q^{(\alpha)}=\epsilon_q$ for all $\alpha$, the eigenvalues
for the Hamiltonian $\mat{H}_{qq}$ of the isolated unit are
$\epsilon_{q}-\ttt$, $\epsilon_{q}+\ttt$, $\epsilon_{q}-\ttt$,
$\epsilon_{q}+\ttt$, $\epsilon_{q}-2\ttt$, $\epsilon_{q}+2\ttt$, while
the corresponding orthonormalized eigenvectors are
\begin{equation} \label{e.benzvecs}
\begin{split}
&\frac{1}{\sqrt{4}}(0,-1,1,-1,1,0)^T,
\frac{1}{\sqrt{4}}(0,1,-1,-1,1,0)^T, \\
&\frac{1}{\sqrt{12}}(-2,-1,-1,1,1,2)^T, 
\frac{1}{\sqrt{12}}(2,-1,-1,-1,-1,2)^T, \\
&\frac{1}{\sqrt{6}}(1,1,1,1,1,1)^T, 
\frac{1}{\sqrt{6}}(-1,1,1,-1,-1,1)^T.
\end{split}
\end{equation}
The first two of the eigenstates have zero weight on the
ring-connecting carbon atoms 1 and 6. Therefore, these eigenstates do
not hybridize with the levels of the adjacent rings and consequently
cannot take part in the transport.  This will be seen explicitly in
the derivation of the propagator.  We note that these results can also
be used to determine a realistic value for the hopping $\ttt$ from the
HOMO-LUMO splitting of benzene.\cite{Pauly07b}

Below we shall only consider the analytically solvable case, where all
onsite energies are set to the same value. We choose this value as our
zero of energy: $\epsilon_{q}^{(\alpha)}=0$ for all $q=1,\ldots,N$ and
$\alpha=1,\ldots,M$.  Later on we shall relax this assumption in order
to describe externally applied dc and ac voltage profiles. In the
absence of such voltages, the inverse propagator [Eq.\
(\ref{e.tridiag})] consists of the blocks 
$\mat{h}_{p,p}=\mat{h}_{0}$, $\mat{h}_{p,p-1}=\mat{h}_{-1}$, and
$\mat{h}_{p,p+1}=\mat{h}_{1}$, where
\begin{equation}\begin{split}
\mat{h}_{0}&=
\left(\begin{matrix}
E_+ & \ttt & \ttt & 0 & 0 & 0 \\
\ttt & E_+ & 0 & \ttt & 0 & 0 \\
\ttt & 0 & E_+ & 0 & \ttt & 0 \\
0 & \ttt & 0 & E_+ & 0 & \ttt \\
0 & 0 & \ttt & 0 & E_+ & \ttt \\
0 & 0 & 0 & \ttt & \ttt & E_+ \\
\end{matrix}\right), \\
\mat{h}_{-1} &=
\left(\begin{matrix}
0 & 0 & 0 & 0 & 0 & \ttu \\
0 & 0 & 0 & 0 & 0 & 0 \\
0 & 0 & 0 & 0 & 0 & 0 \\
0 & 0 & 0 & 0 & 0 & 0 \\
0 & 0 & 0 & 0 & 0 & 0 \\
0 & 0 & 0 & 0 & 0 & 0 \\
\end{matrix}\right)
\end{split}\end{equation}
and $\mat{h}_{1}=[\mat{h}_{-1}]^{T}$. The leads are assumed to couple
only to the terminal carbon atoms, thus making the self-energies
$6\times6$ matrices of the form
\begin{equation}
\mat{\se}_{11}=
\left(\begin{matrix}
\se_L & 0      & \cdots & 0 \\
0        & 0      & \cdots & 0 \\
\vdots   & \vdots & \ddots & 0 \\
0        & 0      & 0      & 0 \\
\end{matrix}\right), \quad
\mat{\se}_{NN}=
\left(\begin{matrix}
0 & 0      & 0      & 0        \\
0 & \ddots & \vdots & \vdots   \\
0 & \cdots & 0      & 0        \\
0 & \cdots & 0      & \se_R \\
\end{matrix}\right).
\end{equation}
We also define the symbol ``tilde'' ($\td{~~}$), which means the
replacement of the first column of a matrix by $\ttu$ followed by
zeros. For example
\begin{equation}
\td{\mat{h}}_{0} =
\left(\begin{matrix}
\ttu & \ttt & \ttt & 0 & 0 & 0 \\
0 & E_+ & 0 & \ttt & 0 & 0 \\
0 & 0 & E_+ & 0 & \ttt & 0 \\
0 & \ttt & 0 & E_+ & 0 & \ttt \\
0 & 0 & \ttt & 0 & E_+ & \ttt \\
0 & 0 & 0 & \ttt & \ttt & E_+ \\
\end{matrix}\right).
\end{equation}

For the evaluation of Eq.\ (\ref{e.genetrans}), we only need the component
$G_{1,MN}=[\mat{G}_{1N}]_{1M}$. Using the
subdeterminants of $\mat{F}=\mat{G}^{-1}$, we have 
\begin{equation} \label{e.detform}
G_{1,MN}=\frac{(-1)^{MN+1}\det[\mat{\igc}(MN|1)]}
{\det[\mat{\igc}]}.
\end{equation}
Here $\mat{O}(i,\ldots,k|j,\ldots,l)$ is the submatrix of $\mat{O}$
obtained by removing the rows $i,\ldots,k$ and columns $j,\ldots,l$.
We shall also denote by $L$ and $R$ the ``leftmost'' and 
``rightmost'' row or column of a matrix. Thus, for example
$\det[\mat{\igc}(MN|1)]=\det[\mat{\igc}(R|L)]$

Let us first concentrate on the denominator of Eq. (\ref{e.detform}).
It is easy to see that $\det[\mat{\igc}]$ can be
written in terms of determinants related to the inverse Green
function $\mat{\igf}=\mat{\gf}^{-1}$ of the uncoupled wire
as follows\cite{Mujica94a}
\begin{equation}
\begin{split}
\det[\mat{\igc}] = \det[\mat{\igf}] &- \se_L\det[\mat{\igf}(L|L)] 
- \se_R\det[\mat{\igf}(R|R)] \\ 
&+ \se_L\se_R\det[\mat{\igf}(L,R|L,R)].
\end{split}
\end{equation}
Furthermore, due to the symmetry of the
molecule, $\det[\mat{\igf}(R|R)]=\det[\mat{\igf}(L|L)]$.  Thus we are
left with calculating three types of determinants. It can be shown
that, for $1<n<N$, all of them satisfy a recursion relation of
the form
\begin{equation}
\begin{split}
\left(\begin{matrix}
D^{(n)} \\
\td D^{(n)}
\end{matrix}\right)
& =
(\Ep^2-\ttt^2)\mat{Y}
\left(\begin{matrix}
D^{(n-1)} \\
\td D^{(n-1)}
\end{matrix}\right) \\
&=
(\Ep^2-\ttt^2)
\left(\begin{matrix}
a & -c \\
c & b \\
\end{matrix}\right)
\left(\begin{matrix}
D^{(n-1)} \\
\td D^{(n-1)}
\end{matrix}\right).
\end{split}
\end{equation}
For example, in the calculation of $\det[\mat{\igf}]$, we have
$D^{(n)}=\det[\mat{\igf}^{(n)}]$ and $\td
D^{(n)}=\det[\td{\mat{\igf}}^{(n)}]$, where the additional superscript
$(n)$ on the matrices denotes the number of the $M\times M$ diagonal
blocks. The elements of the matrix $\mat{Y}$ are given by
\begin{equation}
\begin{split}
a & = (\Ep^2-\ttt^2)(\Ep^2-4\ttt^2) \\
b & = -\ttu^{2}(\Ep^2-\ttt^2)\\
c & = \ttu \Ep(\Ep^2-3\ttt^2).
\end{split}
\end{equation}
Only the initial condition ($n=1$) and the last step of the 
recursion ($n=N$) will differ for the three determinants. 
The recursion relations can be solved by calculating $\mat{Y}^{n}$
explicitly, which can be done by diagonalizing $\mat{Y}$. 
The eigenvalues of $\mat{Y}$ are $\lambda_{1,2}=(a+b\mp\sqrt{(a-b)^2-4c^2})/2$,
while the (unnormalized) eigenvectors are
\begin{equation}
\mat{v}_{1,2}=\left(\frac{a-b\mp\sqrt{(a-b)^2-4c^2}}{2c},1\right)^T.
\end{equation}
Then, if $\mat{V}=(\mat{v}_1, \mat{v}_2)$ and 
$\mat{\Lambda}=\textrm{diag}(\lambda_1,\lambda_2)$,
we have $\mat{Y}^n=\mat{V}\mat{\Lambda}^n\mat{V}^{-1}$.
The result is
\begin{equation}
\mat{Y}^n
=
\left(\begin{matrix}
y^{(n)}_{11} & y^{(n)}_{12}  \\
y^{(n)}_{21} & y^{(n)}_{22}
\end{matrix}\right),
\end{equation}
where the components are given by
\begin{equation}
\begin{split}
y^{(n)}_{11} &= 
\frac{
(\lambda_1^n-\lambda_2^n)(b-a)
+(\lambda_1^n+\lambda_2^n)\sqrt{(a-b)^2-4c^2}}{2\sqrt{(a-b)^2-4c^2}}\\
y^{(n)}_{22} &= 
\frac{
(\lambda_1^n-\lambda_2^n)(a-b)
+(\lambda_1^n+\lambda_2^n)\sqrt{(a-b)^2-4c^2}}{2\sqrt{(a-b)^2-4c^2}}\\
y^{(n)}_{12} &= - y^{(n)}_{21} =
\frac{c(\lambda_1^n-\lambda_2^n)}{\sqrt{(a-b)^2-4c^2}}. \\
\end{split}
\end{equation}
Using these, we can now write explicit expressions for the 
three required determinants. For $\det[\mat{\igf}]$, the recursion
can be started at $n=1$ with the initial conditions $D^{(0)}=1$ and
$\td D^{(0)}=0$ and carried out up to $n=N$. The result is
\begin{equation}
\det[\mat{\igf}^{(N)}] = (\Ep^2-\ttt^2)^N y^{(N)}_{11}.
\end{equation}
The other two determinants require special initial and final steps,
and the results are
\begin{equation}
\begin{split}
\det[\mat{\igf}^{(N)}(L|L)]=&(\Ep^2-\ttt^2)^Ny^{(N)}_{21}/\ttu \\
\det[\mat{\igf}^{(N)}(L,R|L,R)]=&(\Ep^2-\ttt^2)^N[y^{(N-1)}_{21}c \\
                                 &-y^{(N-1)}_{22}b]/\ttu^{2}.
\end{split}
\end{equation}

Next, we consider the determinant in the numerator of Eq.\
(\ref{e.detform}), $\det[\mat{\igc}^{(N)}(R|L)]=\det[\mat{\igf}^{(N)}(R|L)]$. 
It can easily be shown that it satisfies the recursion relation
\begin{equation}
\det[\mat{\igf}^{(N)}(R|L)]=2\ttu \ttt^3(\Ep^2-\ttt^2)\det[\mat{\igf}^{(N-1)}(R|L)]
\end{equation}
and so 
\begin{equation}
\det[\mat{\igf}^{(N)}(R|L)]=2^N(\ttu \ttt^3)^N(\Ep^2-\ttt^2)^N/\ttu.
\end{equation}
Now, the Green function of Eq.\ (\ref{e.detform}) can be 
written as
\begin{equation}
\begin{split}
&G_{1,MN}= \\
&\frac{-(2\ttu \ttt^3)^N/\ttu }{
y^{(N)}_{11}+
\se_{LR}y^{(N)}_{21}/\ttu +
\se_L\se_R\left(y^{(N-1)}_{21}c-y^{(N-1)}_{22}b\right)/{\ttu}^{2}},
\end{split}
\end{equation}
where we used the shorthand $\se_{LR}=\se_L+\se_R$.

It is notable that the common $(\Ep^2-\ttt^2)^N$ factors canceled out
from the final propagator. These factors apparently correspond to the
two eigenvectors of $\mat{h}_0$ [Eq.\ (\ref{e.benzvecs})] 
having zero weight on the ring-connecting
atoms $1$ and $6$. The cancellation is a manifestation of the physical
fact that such localized states cannot contribute to the transport
through the molecule. In the infinite polymer to be discussed below,
these states appear as completely flat bands in the band structure.

To conclude this part, we point out that for $E$ inside the HOMO-LUMO
gap [more precisely, when $(a-b)^2-4c^2>0$] the eigenvalues
$\lambda_{1,2}$ are real-valued and the decay exponent of the
transmission $\tau(E)$ for large $N$ is controlled by the one with a
larger absolute value. Since inside the gap $E\approx 0$, we find that
$\lambda_2>\lambda_1>0$. Then, using Eq.\ (\ref{e.genetrans}) and
omitting $N$-independent prefactors, the decay of the transmission for
large $N$ follows the law
\begin{equation}
\tau(E)\sim 
\left[\frac{\lambda_2(E)}{2\ttu \ttt^3}\right]^{-2N} 
= e^{-2N\ln [\lambda_2(E)/(2\ttu \ttt^3)]}.
\end{equation}
Thus the decay exponent is given by
\begin{equation} \label{e.beta1}
\beta(E)=2\ln [\lambda_2(E)/(2\ttu \ttt^3)].
\end{equation}
We note that for resonant energies, oscillatory dependence of $\tau(E)$
on $N$ can be expected, instead, and for limiting cases
also power-law decay is possible.\cite{Mujica94b} 
Next, we shall reproduce the result for the decay exponent by considering an
infinitely extended polymer.


\subsection{Poly-$p$-phenylene} \label{s.polyphenyl}

\begin{figure}[!tb]
\includegraphics[width=0.85\linewidth,clip=]{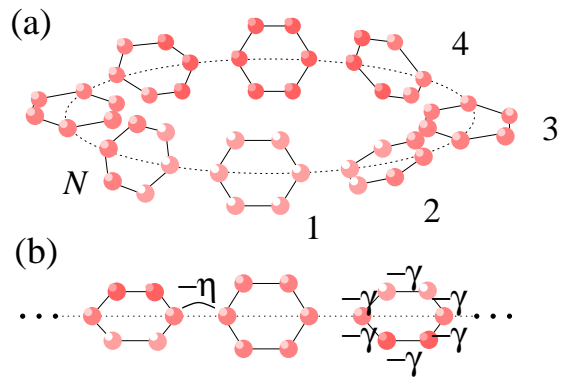}
\caption{(Color online) Phenyl-ring chains: (a) a periodic chain with 
$N$ units and (b) an infinite chain. Case (b) is obtained from (a) in 
the limit $N\rightarrow\infty$. 
}
\label{f.phenylchains}
\end{figure}

For comparison with the ``correct'' evaluation of the propagator and
the decay coefficient for a finite chain, let us consider the
propagator for an infinitely extended polymer. To describe the
polymer, we start from a finite chain with periodic boundary
conditions.  Neglecting curvature effects, the latter actually
represents a ring-shaped oligomer, as depicted in Fig.\
\ref{f.phenylchains}(a).

Let us first consider the eigenstates of the periodic chain.
The Hamiltonian
$H^{(\alpha,\beta)}_{pq}=\langle\chi^{(\alpha)}_p|\hat{H}|\chi^{(\beta)}_q\rangle$
is of the general form
\begin{equation}
\mat{H}=\left(\begin{matrix}
\mat{H}_{0}  & \mat{H}_{1}  &        &        & \mat{H}_{-1} \\
\mat{H}_{-1} & \mat{H}_{0}  & \mat{H}_{1}  &        &        \\
       & \ddots & \ddots & \ddots &        \\
       &        & \mat{H}_{-1} & \mat{H}_{0}  & \mat{H}_{1}  \\
\mat{H}_{1}  &        &        & \mat{H}_{-1} & \mat{H}_{0}  \\
\end{matrix}\right),
\end{equation}
where $\mat{H}_{0,\pm1}$ are the $M\times M$ matrices ($M=6$) of Eqs.\
(\ref{e.diagblock}) and (\ref{e.offdiagblock}), with
$\epsilon_{q}^{(\alpha)}=0$. (Again, only nonzero elements are
indicated.)  The normalized eigenvectors $\mat{\psi}^{(n)}_{p}(k)$
satisfying
\begin{equation}
\sum_{q}\mat{H}_{pq}\mat{\psi}^{(n)}_q(k) = E^{(n)}(k)\mat{\psi}^{(n)}_p(k)
\end{equation}
are of the Bloch form
$\mat{\psi}^{(n)}_{q}(k) = e^{\iu k q d} \mat{\phi}^{(n)}(k)/\sqrt{N}$,
where $\mat{\phi}^{(n)}(k)$ are the normalized eigenvectors of
\begin{equation}
\mat{H}(k) = e^{ikd}\mat{H}_{1} + \mat{H}_0 + e^{-ikd}\mat{H}_{-1}
\end{equation}
with the eigenvalue $E^{(n)}(k)$, and $n=1,\ldots,M$.  Due to the
finiteness of the wire, the $k$ values are restricted to
$k_{\mu}=2\pi\mu/Nd$, where $\mu$ is an integer and $d$ is the lattice
constant (the length of a single phenyl-ring unit).

The spectral decomposition of the (retarded) propagator 
$\mat{\gb}(E)=(\Ep\mat{1}-\mat{H})^{-1}$ of the chain is of the form
\begin{equation} \label{e.ringprop}
\gb^{(\alpha,\beta)}_{pq}(E)=\sum_{\mu,n}
\frac{\langle\chi^{(\alpha)}_{p}|\psi^{(n)}(k_\mu)\rangle
\langle\psi^{(n)}(k_\mu)|\chi^{(\beta)}_q\rangle}
{E_{+} - E^{(n)}(k_\mu)},
\end{equation}
with the Bloch states
\begin{equation}
|\psi^{(n)}(k_\mu)\rangle = \frac{1}{\sqrt{N}}
\sum_{p=-\lceil N/2\rceil+1}^{\lfloor N/2\rfloor}
e^{\iu k_\mu p d} \sum_{\alpha=1}^{M}\phi^{(n)}_\alpha(k_\mu)
|\chi^{(\alpha)}_p\rangle.
\end{equation}
In the limit of large $N$ [Fig.\ \ref{f.phenylchains}(b)], we can 
use
$N^{-1}\sum_\mu\rightarrow(d/2\pi)\int_{-\pi/d}^{\pi/d}\upd k$
to turn the summation into an integral over the first Brillouin zone.
In this case, there are $M=6$ bands with energies
\begin{equation} \label{e.bands}
\begin{split}
E^{(1,2)}(k) &= \pm \ttt \\
E^{(3,4)}(k) &= \pm \frac{1}{\sqrt{2}}
\sqrt{{\ttu}^{2}+5\ttt^2 - 2B(k)} \\
E^{(5,6)}(k) &= \pm \frac{1}{\sqrt{2}}
\sqrt{{\ttu}^{2}+5\ttt^2 + 2B(k)},
\end{split}
\end{equation}
where 
\begin{equation}
\begin{split}
B(k) &= \frac{1}{2}\sqrt{({\ttu}^{2}+3\ttt^2)^2+16\ttu \ttt^3\cos(kd)}.
\end{split}
\end{equation}
Clearly we have the symmetries $E^{(1)}(k) = -E^{(2)}(k)$, $E^{(3)}(k)
= -E^{(4)}(k)$, and $E^{(5)}(k) = -E^{(6)}(k)$.  For $n=1,2$ the bands
are completely flat, and the corresponding eigenvectors
$\mat{\phi}^{(1,2)}(k)$ are as in Eq.\ (\ref{e.benzvecs}), i.e.,
independent of $k$ and completely localized on atoms $\alpha=2,3,4,5$.
Thus for $p\neq q$, they do not contribute to the propagator in Eq.\
(\ref{e.ringprop}).  For $n=3,4,5,6$, the vectors are very
complicated, but they are not needed in the following.

To compare with the result of Sec.\ \ref{s.oligophenyl}, we should now
calculate, for example, the component $g_{pq}^{(1,6)}$. However,
expecting the decay exponent to be independent of $\alpha$ and
$\beta$, we consider the simpler case $\Tr[\mat{\gb}_{pq}]=\sum_\alpha
\gb^{(\alpha,\alpha)}_{pq}$.  Due to the orthonormality
$\sum_{\alpha}\phi^{(m)}_{\alpha}\phi^{(n)*}_{\alpha}=\delta_{mn}$,
the dependence on the vector components then drops out. Thus, for
$p\neq q$
\begin{equation}
\begin{split}
\sum_\alpha g^{(\alpha,\alpha)}_{pq} & =
4EA
\frac{d}{2\pi}\int_{-\pi/d}^{\pi/d}\upd k 
\frac{e^{\iu k d (p-q)}}{A^2-B^2(k)}, \\
\end{split}
\end{equation}
where we defined
\begin{equation}
\begin{split}
A & = \Ep^2-\frac{1}{2}({\ttu}^{2}+5\ttt^2),
\end{split}
\end{equation}
such that $\Ep^2-[\epsilon^{(3,5)}(k)]^2= A \pm B(k)$.
Defining now $z=e^{\iu k d}$, the integral can be turned into 
a contour integral around the contour $|z|=1$
\begin{equation}
\begin{split}
\sum_\alpha g^{(\alpha,\alpha)}_{pq} & =
-\frac{2EA}{2\pi\iu \ttu \ttt^3}
\oint_{|z|=1}\upd z
\frac{z^{p-q}}{(z-z_+)(z-z_-)}, \\
\end{split}
\end{equation}
where the poles $z_\pm$ are determined from the 
equation $z^2-[4A^2-({\ttu}^{2}+3\ttt^2)^2](8\ttu \ttt^3)^{-1}z+1=0$.
They are given by
\begin{equation}
z_\pm = \frac{4A^2-({\ttu}^{2}+3\ttt^2)^2}{16\ttu \ttt^3} \pm
\sqrt{\left[\frac{4A^2-({\ttu}^{2}+3\ttt^2)^2}{16\ttu \ttt^3}\right]^2-1}
\end{equation}
such that $z_+=1/z_-$, and we choose the signs so that $z_-$ is inside
the contour $|z|=1$. In addition to this, assuming that $p<q$, there
is a pole of order $q-p$ at $z=0$.  The integral can then be evaluated
using residue techniques, with the result
\begin{equation}
\sum_\alpha g^{(\alpha,\alpha)}_{pq}
=\frac{2EA}{\ttu \ttt^3}\frac{z_+^{p-q}}{z_+-z_-}.
\end{equation}
This leads to an exponential decay of the propagator with growing
$q-p>0$, when $E$ is off-resonant (in which case $z_{\pm}$ are
real-valued). Using this result, we can give an estimate for the decay
of the transmission function [Eq.\ (\ref{e.genetrans})] through a
finite chain of length $N$ by replacing $G_{1,MN}$ with
$\Tr[\mat{g}_{1N}]/M$. This yields
\begin{equation}
\tau(E)\sim [z_+(E)]^{-2N} = e^{-2N\ln [z_+(E)]},
\end{equation}
and thus the exponent
\begin{equation} \label{e.beta2}
\beta(E) = 2\ln [z_+(E)].
\end{equation}
It can be checked that this result is, in fact, equal to the
result [Eq.\ (\ref{e.beta1})] obtained for the finite chain.

It is thus seen explicitly that the decay coefficient of the
off-resonant transmission does not in any way depend on the coupling
of the molecule to the leads. It should be kept in mind, however, that
the relative position of $E_F$ within the HOMO-LUMO gap depends on the
electrode-lead coupling and the charge transfer effects. This
information is still needed for predicting the decay exponent
$\beta(E_F)$ of the conductance.

The analytical results presented in this and the previous section can
be used for understanding the behavior of the transmission function
upon changes in the parameters.  For example, it should be noted that
when $\ttu $ is made smaller, the band gap around $E\approx 0$ becomes
larger, and at the same time the decay exponent $\beta(E)$ grows. In
this way, the conductance of a molecular junction can be controlled,
for example, by introducing side groups to control the tilt angles
$\varphi$ between the phenyl rings.\cite{Pauly07a,Pauly07b}


\section{Physical observables and numerical results} \label{s.results}

In this Section we present numerical results based on our
model. Throughout, we employ the ``wide-band'' approximation for the
lead self-energies, such that $\se_L(E)=-\iu\ga_L/2$ and
$\se_R(E)=-\iu\ga_R/2$, with energy-independent constants $\ga_{L,R}$ .
Furthermore we only consider the symmetric case $\ga_L=\ga_R=\ga$.
First we briefly describe how we generalize the theory, as presented
above, to take into account static and time-dependent voltage
profiles.  Then we concentrate on near-equilibrium (or
``linear-response'') properties, using as examples the conductance,
the thermopower, and the conductance enhancement due to light with
low intensity and frequency.  In this case, knowledge of the zero-bias
transmission function calculated above is sufficient, and we can
discuss the length dependence of the transport properties in a simple
way. After that we consider the dc current in the presence of an ac
driving field of more general amplitude and frequency, first
concentrating on the case of infinitesimal dc bias, and finally on the
$I$-$V$ characteristics.

\begin{figure}[!tb]
\includegraphics[width=0.85\linewidth,clip=]{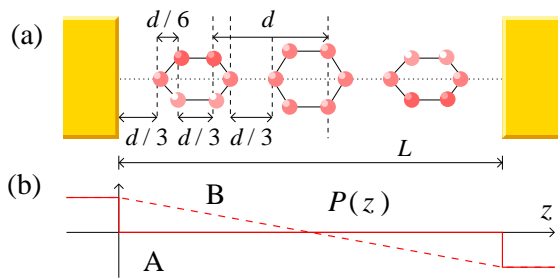}
\caption{(Color online) (a) The coordinates of the carbon atoms
in the direction $z$ along the molecular wire. The left electrode is
at $z=0$ and the length of a phenyl-ring unit is $d$. (b) Relative
variation of the onsite energies for two different voltage profiles, A
and B. The profile function $P(z)$ describes how the harmonic voltage
$V(t)=V+V_{ac}\cos(\omega t)$ is assumed to drop over the junction,
the voltage at $z$ being given by $V(z,t)=V(t)P(z)$.  }
\label{f.profiles}
\end{figure}

\subsection{Voltage profiles} \label{s.profiles}

When considering finite dc or ac biases within a non-selfconsistent TB
model that cannot account for screening effects, one of the obvious problems
is how to choose the voltage profile. Throughout the discussion, we
shall refer to two possible choices, as depicted in Fig.\ \ref{f.profiles}.
They are in some sense limiting cases, and the physically most
reasonable choice should lie somewhere in between.  Profile A
assumes the external electric fields to be completely screened inside
the molecule, such that the onsite energies are not modified,
while B corresponds to the complete absence of such screening. In
both cases, we can write the time-dependent onsite energies as
$\epsilon_p^{(\alpha)}(t)=eV(t)P(z_p^{(\alpha)})$, where
$z_p^{(\alpha)}$ are the distances of the carbon atoms from the left
metal surface, and $V(t)=V+V_{ac}\cos(\omega t)$.  In case A, $P(z)=0$
inside the junction, while in case B $P(z)=(L-2z)/(2L)$, where
$L=Nd+d/3$ is the distance between the two metal surfaces.

The profile B is more complicated, because the voltage ramp breaks the
homogeneity of the wire. In this case the current must be calculated
with the method outlined in App.\ \ref{s.lightcurr}.  In the case of
profile A, however, the $I$-$V$ characteristics can be calculated
based on the knowledge of the zero-bias transmission function in the
absence of light, $\tau(E)$. As discussed in App.\ \ref{s.simplecurr},
the current is given by\cite{Pedersen98,Kohler04,Platero04}
\begin{equation} \label{e.simplec}
\begin{split}
I(V;\alpha,\omega)=&
\frac{2e}{h}
\sum_{l=-\infty}^\infty 
\left[J_{l}\left(\frac{\alpha}{2}\right)\right]^2
\int \upd E 
\tau(E+l\hbar\omega) \\  
&\times [f_L(E)-f_R(E)].
\end{split}
\end{equation}
The low-temperature zero-bias conductance
then takes the particularly simple form\cite{Viljas07a,Viljas07b}
\begin{equation} \label{e.photoc}
G_{dc}(\alpha,\omega)=G_0\sum_{l=-\infty}^{\infty}
\left[J_l\left(\frac{\alpha}{2}\right)\right]^2\tau(E_F+l\hbar\omega).
\end{equation}
Here $l$ indexes the number of absorbed or emitted photons, $J_l(x)$
is a Bessel function of the first kind (of order $l$), and
$\alpha=eV_{ac}/\hbar\omega$ is the dimensionless parameter describing
the strength of the ac drive. Note that
$G_{dc}(\alpha,\omega=0)=G_{dc}(\alpha=0,\omega)=G_0\tau(E_F)=G_{dc}$.
Equation (\ref{e.simplec}) may equally well be written in the
form\cite{Tien63,Tucker85}
\begin{equation} \label{e.simplec2}
I(V;\alpha,\omega)=
\sum_{l=-\infty}^{\infty}
\left[J_l\left(\frac{\alpha}{2}\right)\right]^2
I_{0}(V+2l\hbar\omega/e),
\end{equation}
where $I_{0}(V)$ is the $I$-$V$ characteristic in the absence of
light [Eq.\ (\ref{e.curr})]. Below, the results from these 
formulas are compared to the numerical results for profile B.

\begin{figure}[!tb]
\includegraphics[width=0.85\linewidth,clip=]{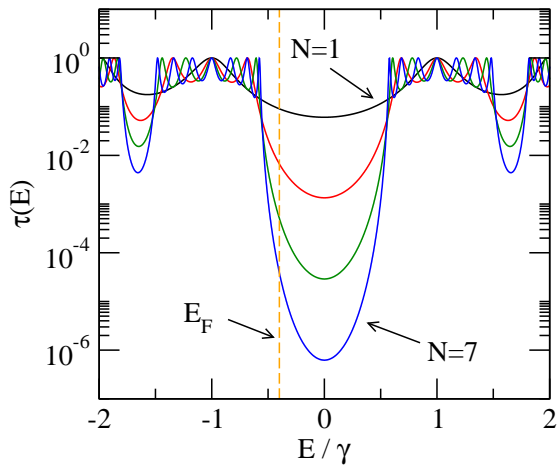}
\caption{(Color online) Transmission functions for the oligophenylene wires with
lengths $N=1,3,5,7$. 
The parameters are $\ga/\ttt=0.5$, $\varphi=40^\circ$, $E_F/\ttt=-0.4$,
as discussed in the text.
}
\label{f.trans}
\end{figure}

In Fig.\ \ref{f.trans} we plot the zero-bias transmission functions
for wires with $N$ between $1$ and $7$. Notice that the four energy
bands numbered 3-6 in Eq.\ (\ref{e.bands}) are all visible, being
separated by the HOMO-LUMO gap at $E/\ttt\approx 0$ and the additional
gaps at $E/\ttt\approx\pm 1.7$.  Here we use the parameters
$\ga/\ttt=5.0$, $\varphi=40^\circ$ (i.e. $\eta/\ttt\approx0.77$),
and set the Fermi energy to $E_F/\ttt=-0.4$.  These values are close
to those used in Ref.\ \onlinecite{Pauly07b}, where they were
extracted from a fit to results for
gold-oligophenylene-gold contacts based on density-functional theory
(DFT). We shall continue to use them everywhere below.  A DFT
calculation for the HOMO-LUMO splitting of benzene, together with the
results preceding Eq.\ (\ref{e.benzvecs}), yields the hopping
$\ttt\approx 3$ eV. The length of a phenyl-ring unit is
approximately $d=0.44$ nm, and the largest ac electric fields
$V_{ac}/L$ considered will be on the order of $10^9$ V/m.
The photon energies $\hbar\omega$ will mainly be kept 
below the energy of the HOMO-LUMO gap of the oligophenylene.

\begin{figure}[!tb]
\includegraphics[width=0.85\linewidth,clip=]{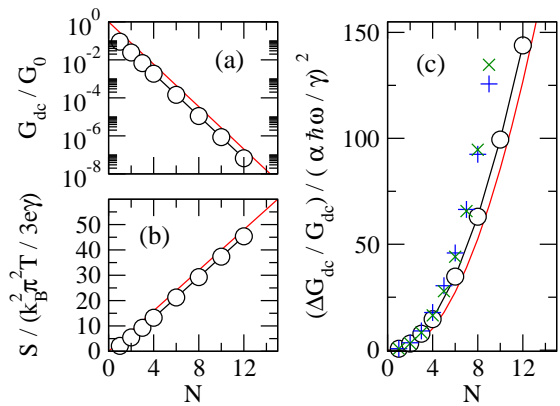}
\caption{(Color online)
Dependence of observables on the number of units $N$: (a) Conductance,
(b) Seebeck coefficient, and (c) the light-induced relative
conductance-enhancement.  The circles correspond to values extracted
from the $\tau(E)$ function [Fig.\ \ref{f.trans}], using Eqs.\
(\ref{e.conductance}), (\ref{e.thermopower}), and
(\ref{e.correction}).  The red lines correspond to the simple order-of
magnitude estimates of Eqs.\ (\ref{e.explaw}), (\ref{e.linlaw}), and
(\ref{e.quadlaw}), with the analytically calculated $\beta(E)$. In (c)
the crosses ($\times$ for profile A and $+$ for profile B) show
numerical results with the finite values $\alpha=0.5$ and
$\hbar\omega/\ttt=0.05$ (see Sec.\ \ref{s.zerobias}).  } \label{f.anal}
\end{figure}
%


\subsection{Near-equilibrium properties} \label{s.neareq}

Let us start by illustrating the usefulness of the analytical results
of Sec.\ \ref{s.phenyl} with a few examples. We concentrate on low
temperatures and small deviations from equilibrium. In addition to the
linear-response conductance
\begin{equation} \label{e.conductance}
G_{dc}=G_0\tau(E_F),
\end{equation}
we shall consider the thermopower, or Seebeck coefficient.
At low enough temperature $T$, this is given
in terms of the zero-bias transmission function $\tau(E)$ 
as\cite{Sivan86,Paulsson03,Zheng04,Segal05}
\begin{equation} \label{e.thermopower}
S=-\frac{\pi^2k_B^2 T}{3e}
\frac{\tau'(E_F)}{\tau(E_F)},
\end{equation}
where prime denotes a derivative.  Thus it measures the
logarithmic \emph{first derivative} of the transmission function at
$E=E_F$. The sign of this quantity carries information about the
location of the Fermi energy within the HOMO-LUMO gap of molecular
junction.\cite{Paulsson03} The third quantity we shall consider is the
photoconductance.  In the limit $\alpha\ll 1$ and $\hbar\omega/\ttt\ll 1$
we can expand $\tau(E)$ and the Bessel functions in Eq.\
(\ref{e.photoc}) (see App.\ \ref{s.simplecurr}) to leading order in
these small quantities, yielding $G_{dc}(\omega)=G_0\tau(E_F) +
G_0(\alpha\hbar\omega)^2\tau''(E_F)/16$.  Defining then the
light-induced conductance correction
$\Delta G_{dc}(\omega)=G_{dc}(\omega)-G_{dc}(\omega=0)$, where
$G_{dc}(\omega=0)=G_{dc}=G_0\tau(E_F)$, the relative correction
becomes
\begin{equation} \label{e.correction}
\frac{\Delta G_{dc}(\alpha,\omega)}{G_{dc}}
=\frac{(\alpha\hbar\omega)^2}{16}\frac{\tau''(E_F)}{\tau(E_F)}.
\end{equation}
We thus see that this quantity gives experimental access to the
\emph{second derivative} of the transmission function at $E=E_F$.
Note that in this approximation, which can be seen as an adiabatic or
``classical'' limit,\cite{Tucker85} the conductance correction depends only
on the driving field through the ac amplitude
$V_{ac}=\alpha\hbar\omega/e$.

As discussed above, it is reasonable to assume that for large enough
$N$, the transmission function $\tau(E)$ satisfies the exponential
decay law
\begin{equation} \label{e.explaw}
\tau(E)\sim C(E)e^{-\beta(E)N} 
\end{equation}
at the off-resonant energies $E\approx E_F$. Let us furthermore
assume that $C(E)$ is only weakly $E$-dependent.
Then it is clear that the Seebeck coefficient will have the following 
simple linear dependence on $N$ (Refs.\ \onlinecite{Segal05,Pauly07b}):
\begin{equation} \label{e.linlaw}
S\propto\tau'(E_F)/\tau(E_F)\sim-\beta'(E_F)N.
\end{equation}
In contrast, the light-induced conductance correction satisfies a
quadratic law
\begin{equation} \label{e.quadlaw}
\begin{split}
\Delta G_{dc}(\omega)/G_{dc}
&\propto\tau''(E_F)/\tau(E_F) \\
&\sim-\beta''(E_F)N+[\beta'(E_F)]^2N^2.
\end{split}
\end{equation}
Deviations from these laws can follow from the 
energy-dependence of $C(E)$.

In Fig.\ \ref{f.anal} we demonstrate these length dependences within
our model for the oligophenylene junctions. The circles connected by
lines show the results based on the transmission functions of Fig.\
\ref{f.trans}, using Eqs.\ (\ref{e.conductance}),
(\ref{e.thermopower}), and (\ref{e.correction}).  The separate solid
lines are the estimates of Eqs.\ (\ref{e.explaw}), (\ref{e.linlaw}),
and (\ref{e.quadlaw}), based on the analytic result for
$\beta(E)$. The result for $\Delta G_{dc}(\omega)/G_{dc}$ is
furthermore compared with some example results for finite $\alpha$ and
$\omega$, using $\alpha=0.5$ and $\hbar\omega/\ttt=0.05$ (see below).
Although Eq.\ (\ref{e.correction}) was derived above by assuming the
profile A, the result appears to be rather well satisfied for profile
B as well.


%
\begin{figure}[!tb]
\includegraphics[width=0.85\linewidth,clip=]{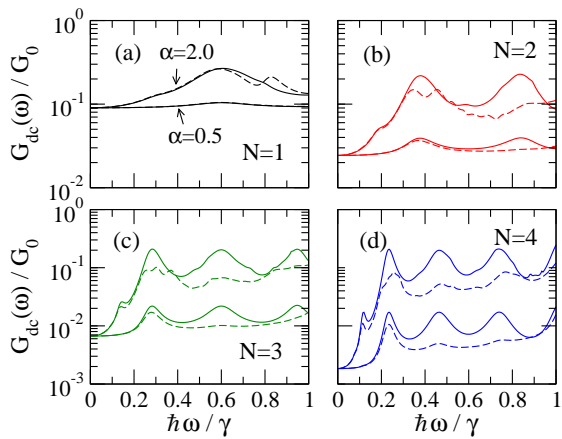}
\caption{(Color online) Zero-bias conductance for different driving frequencies $\omega$ 
and driving strengths $\alpha=eV_{ac}/\hbar\omega$. 
Panels (a)-(d) are for $N=1,\ldots,4$. The solid lines correspond to profile A,
and the dashed lines to profile B. The lower pair of curves is for 
$\alpha=0.5$, and the upper pair for $\alpha=2.0$.
}
\label{f.photoc}
\end{figure}
\begin{figure}[!tb]
\includegraphics[width=0.85\linewidth,clip=]{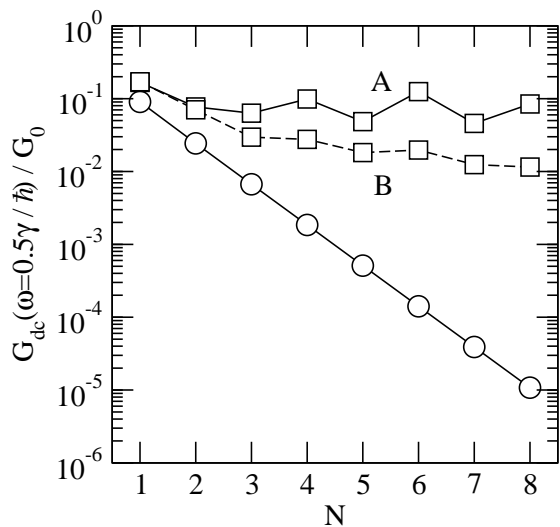}
\caption{(Color online) Dependence of the conductance on $N$. 
Circles represent the conductance in the absence of light, while the
squares are for light with $\hbar\omega/\ttt=0.5$ and
$\alpha=1.5$. The solid line is for profile A and the dashed line for
profile B.  }
\label{f.decay}
\end{figure}

\subsection{Zero-bias conductance at finite drive frequencies and
amplitudes} \label{s.zerobias}

Next we consider the zero-bias photoconductance $G_{dc}(\omega)$ for
light whose frequencies and intensities are not restricted to the
adiabatic limit. We have discussed this case previously, based on DFT
results for gold-oligophenylene-gold contacts.\cite{Viljas07b} There,
however, the analysis was based solely on the simple formula of Eq.\
(\ref{e.photoc}). Here we show that those results are not expected to
change in an essential way within a more refined theory, since the
results of our TB model are not very different for the two voltage
profiles A and B.  This is seen in Fig.\ \ref{f.photoc}, where we show
$G_{dc}(\omega)$ for $N=1,\ldots,4$ as a function of $\omega$ for two
values of $\alpha$, and for both profiles. The results for profile A
again follow from Eq.\ (\ref{e.photoc}), but the results for B require
a more demanding numerical calculation (see App.\
\ref{s.lightcurr}). In both cases the effect of light is to increase
the conductance considerably. The physical reason is that the
photoassisted processes, where electrons emit or absorb radiation
quanta, brings the electrons to energies outside of the HOMO-LUMO gap,
where the transmission probability is higher.  This happens when
$\hbar\omega$ exceeds the energy difference between the Fermi energy
and the closest molecular orbital, in this case the HOMO.  The main
difference between the two profiles is that in case B, the sharp
resonances at some frequencies are smeared out, and thus the
light-induced conductance enhancement tends to be smaller. The
increase can still be an order of magnitude or more. 

The dependence of this effect on the length of the molecule is
still illustrated in Fig.\ \ref{f.decay}, where the conductances in the
absence of light and in the presence of light with
$\hbar\omega/\ttt=0.5$ and $\alpha=1.5$ are shown as a function of
$N$. While the conductance in the absence of light has a strong
exponential decay, in the presence of light this decay is much slower.
For profile A the conductance actually oscillates periodically, while
in the case of profile B the oscillations are superimposed on a
background of slow exponential decay. In the DFT-based
results\cite{Viljas07b} the oscillations were not present, or at least
not visible for the cases $N=1,\ldots,4$ considered there.  Indeed,
they are likely to be artifacts of the our TB model that neglects all
other than $\pi$-orbital contributions, as well as uses the wide-band
approximation.

The results of Fig.\ \ref{f.decay} can also be stated in terms of the
relative conductance-enhancement $\Delta G_{dc}(\omega)/G_{dc}$. For
large $\alpha$ and $\omega$, the increase of this quantity with $N$ is
exponential for both profiles A and B. This should be contrasted with
the quadratic behavior for small $\alpha$ and $\omega$ 
[Eq.\ (\ref{e.quadlaw})].  Thus, the fact
that the results indicated by the crosses in Fig.\ \ref{f.anal} exceed
the result of Eq.\ (\ref{e.correction}) is understandable.


%
\begin{figure}[!tb]
\includegraphics[width=0.85\linewidth,clip=]{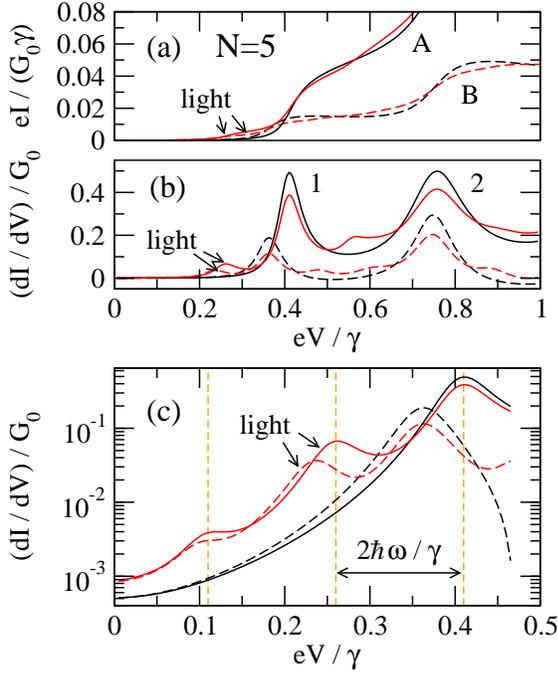}
\caption{(Color online) (a) $I$-$V$ characteristics ($N=5$) 
with and without light for profiles A (solid lines) and B (dashed
lines). Results in the presence of light with $\alpha=1.5$ and
$\hbar\omega/\ttt=0.075$ are indicated with an arrow.  (b) The
corresponding differential conductances.  (c) Same as (b), but
concentrating on the low-bias regime and on a logarithmic scale. The
vertical dashed lines indicate the approximate positions of the main
peak and the light-induced side peaks.  They are all separated by
$2\hbar\omega/e$ in voltage.  }
\label{f.diffc}
\end{figure}
\begin{figure}[!tb]
\includegraphics[width=0.85\linewidth,clip=]{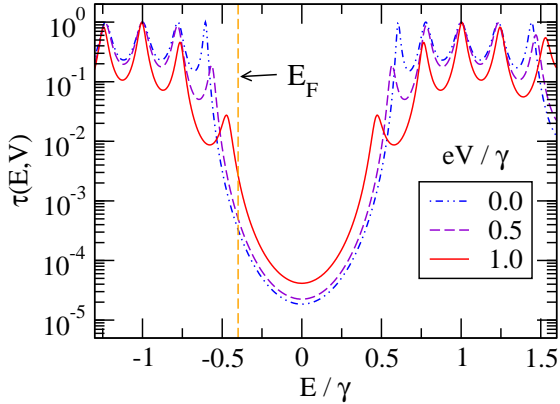}
\caption{(Color online)
The voltage-dependent transmission function at three voltages
for the wire with $N=5$ and profile B. For profile A the result
is independent of voltage and equal to $\tau(E,V=0)$.
\label{f.vtrans}
}
\end{figure}

\subsection{Current-voltage characteristics} \label{s.iv}

Finally we discuss the effects of light at finite voltages $V$.  Let
us first consider the properties of the $I$-$V$ characteristics in the
absence of light. Examples are shown in Fig.\ \ref{f.diffc}(a) for the
case $N=5$. They consist of consecutive steps,\cite{Elbing05} which
appear every time a new molecular level comes into the bias window
between $\mu_L$ and $\mu_R$. These steps are seen as the peaks 1 and 2
in the differential conductance $dI/dV$ shown in Fig.\
\ref{f.diffc}(b).  The first one occurs roughly at the voltage
$V_{1}=2(E_F-E_{HOMO})/e$, where $E_{HOMO}$ is the energy of the
HOMO. The factor $2$ arises from the symmetric division of the
voltages with respect to the molecular energy levels.  In the case of
profile B, the currents tend to be smaller than for profile A, but the
current steps occur at roughly the same voltages. It should also be
noticed that for profile B, a small negative differential conductance
is present following some of the steps. The origin of this is the
localization of the molecular eigenstates due to the dc voltage ramp,
which suppresses the transmission resonances.\cite{Mujica96} This can
be seen in the voltage-dependent transmission functions $\tau(E,V)$ in
Fig.\ \ref{f.vtrans}.

In the presence of light, the step structure of the $I$-$V$ curves is
modified.  For profile A, the results follow simply from Eq.\
(\ref{e.simplec}) or (\ref{e.simplec2}), but for profile B a fully
numerical treatment is again needed. In Fig.\
\ref{f.diffc} the results for $\alpha=1.5$ and $\hbar\omega/\ttt=0.075$ 
are shown as the curves indicated with arrows. In Fig.\
\ref{f.diffc}(a) it is seen that the current for voltages below the
steps is increased, and decreased above them. This removes the
negative differential conductance present in the case of profile
B. These changes are associated with the appearance of additional
current steps. Here we concentrate only on the additional steps in the
low-bias regime at voltages $V\lesssim V_{1}$, as the relative changes
are largest there. Fig.\ \ref{f.diffc}(c) shows the differential
conductance on a logarithmic scale in this voltage region.  It can be
seen that there are multiple extra peaks below the main peak, all of
which are separated by voltages $2\hbar\omega/e$ from each other.
These peaks are ``images'' of the main peak at $V=V_{1}$, and are
easily understood based on Eq.\ (\ref{e.simplec2}).  For profile B all
the peaks are moved to slightly smaller voltages and their spacing is
reduced, since finite voltages tend to also suppress the transmission
gap (see again Fig.\ \ref{f.vtrans}). Notice that, in contrast to
high dc biases [Fig.\ \ref{f.diffc}(a,b)], in the low-bias regime
[Fig.\ \ref{f.diffc}(c)] the results depend only weakly on the choice
of the voltage profile. Thus the predictions of the model appear to be
robust.  To observe the side steps, the radiation frequency should be
large enough such that the steps are not ``lost'' under the broadening
of the main steps. On the other hand, it should be small enough to have
at least one step present. Thus, if the voltage broadening of the main
step at $V=V_1$ is approximately $\Delta_1/e$, then we require
$\Delta_1\lesssim\hbar\omega<E_F-E_{HOMO}$.

\begin{figure}[!tb]
\includegraphics[width=0.85\linewidth,clip=]{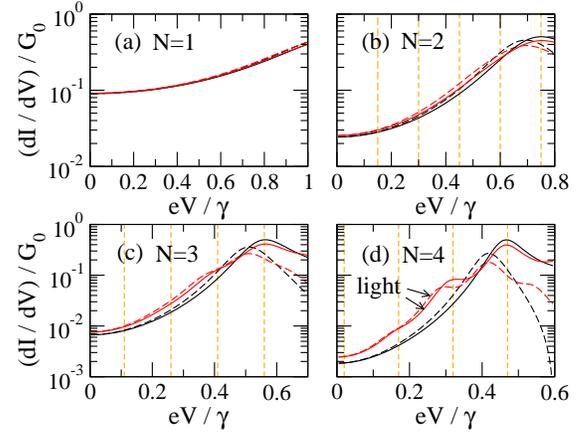}
\caption{(Color online) Same as Fig.\ \ref{f.diffc}(c) but for wires
with $N=1,\ldots,4$.
\label{f.diffc2}
}
\end{figure}
Figure \ref{f.diffc2} additionally shows the low-bias differential
conductances for $N=1,\ldots,4$, with other parameters chosen as in
Fig.\ \ref{f.diffc}(c).  It is seen that the effects of light quickly
become weaker, as the length of the molecule decreases. In the case
$N=4$, small side peaks are still observed. Larger effects could be
obtained by increasing the parameter $\alpha$.

Similar-looking additional steps are visible in the $I$-$V$
characteristics of an extended-H\"uckel model for xylyl-dithiol in
Ref.\ \onlinecite{Tikhonov02b}. Despite the differences in magnitudes
of parameters, and slight asymmetries in the geometries, it is likely
that some of those steps have essentially the same origin as explained
above. However, the most striking result in that reference was the
overall order-of-magnitude increase in the current.


\section{Conclusions and discussion} \label{s.discussion}

In this paper we have studied a $\pi$-orbital tight-binding model to
describe elastic and photoassisted transport through
metal-molecule-metal contacts based on oligophenylenes. In contrast
with simpler linear chain models that have previously been studied in
great detail, our model describes a specific molecule, and its
parameters can be directly associated with quantities obtainable from
DFT simulations, for example.  Models of this type can be of value in
analyzing the results of more detailed \emph{ab-initio} or DFT
calculations,\cite{Pauly07b} and in making at least qualitative
predictions in situations where such calculations would be
prohibitively costly.

We first showed that at zero voltage bias the model can be studied
analytically in a similar fashion as the simpler linear chain models.
In particular, we derived an expression for the decay exponent of the
off-resonant transmission function.  We then discussed the length
dependence of the dc conductance, the thermopower, and the relative
light-induced conductance enhancement in the case of light with a low
intensity ($\alpha$) and low frequency ($\omega$). The conductance
enhancement was found to scale quadratically with length.  For large
$\alpha$ and $\omega$, the relative enhancement increases
exponentially with length.  Finally it was shown, by numerical
calculations, that the current-voltage characteristics are modified in
the presence of light by the appearance of side steps with a voltage
spacing $2\hbar\omega/e$. We demonstrated that the predictions of the
model are robust with respect to variations in the assumed voltage
profiles. This provides further support for our previous results on
the photoconductance.\cite{Viljas07b}

In our work, only symmetrical junctions with symmetrical voltage
profiles were studied. Asymmetries can modify our results through the
introduction of rectification effects,\cite{Grafstrom02} and can
change the positions of the light-induced current steps.  The
experimental observation of additional steps with a spacing related to
the frequency of the light would nevertheless provide more compelling
evidence for the presence of photoassisted transport than a
conductance enhancement alone. The latter can also have other
causes.\cite{Guhr07a}

We note that the light-induced current steps are similar to the steps
observed in current-voltage characteristics of microwave-irradiated
superconducting tunnel junctions, where they result from photoassisted
quasiparticle tunneling.\cite{Tinkham,Tien63} In that case, the main
difference is that the energy gap necessary for the effect is located
in the macroscopic electrodes, while the transmission through the
tunnel barrier depends only weakly on energy and voltage. As a result,
the current steps have a voltage spacing of precisely $\hbar\omega/e$.
These effects are exploited in the detection of microwaves in
radioastronomy.\cite{Tucker85} Similarly, one may imagine properly
engineered molecular contacts as detectors of light in the infrared or
visible frequency range.

In terms of our model, to increase the chances of observing the
light-induced current steps, the aim should be to minimize the
broadening $\Delta_1/e$ of the first main current step at voltage
$V_1$, and to maximize $\alpha$. Also, a wire with a large enough
$V_1$ should be used. The broadening $\Delta_1$ is related to the
sharpness of the transmission resonances, and thus to the length of
the molecule and its coupling to the electrodes, described by
$\Gamma$.  A decrease of $\Gamma$, however, increases the importance
of Coulomb correlations. Their effect on photoassisted transport has
recently been discussed within simple models.\cite{Kaiser06,Li07}
Increase of $\alpha$ through the light intensity, in turn, increases
the heating of the electrodes\cite{Guhr07a} and the excitation of
local molecular vibrations.\cite{Viljas05} These may affect the
geometry through thermal expansion\cite{Grafstrom02} and structural
deformations, but will also give rise to an incoherent component to
the current.\cite{Segal02} At high enough photon energies, also the
direct excitation of electrons on the molecule may become important.
The relaxation of such excitations due to various mechanisms (creation
of electron-hole pairs in the electrodes, spontaneous light emission)
should thus also be considered.\cite{Galperin05} Also conformational
changes of the molecule are possible.\cite{vanderMolen06} Finally, a
proper treatment of screening effects on the molecule and in the
electrodes, the excitation of plasmons, and their role in the field
enhancement\cite{Grafstrom02} are other issues that should be studied
in more detail.

Of course, for the investigation of most of these issues,
noninteracting models of the type presented above are not sufficient.
Strong time-dependent electric fields may have effects that can only
be captured by self-consistent theories taking properly into
account the electron correlations due to Coulomb interactions. These
interactions may influence the electronic structure in a way that
would, at least, require the parameters of our model to be readjusted
in the presence of the light. Even the geometry of the junction can
become unstable, and so it should in principle be optimized with the
light-induced effects included. Time-dependent density-functional
theory is showing some promise for the treatment of such
problems.\cite{Kurth05,Galperin07} In addition to DFT, also more
advanced computational schemes are being developed to handle
correlation effects.\cite{Dahnovsky05,Thygesen07} A systematic
investigation of the optical response of metal-molecule-metal
contacts, and thus the testing of the predictions of the simple
models,\cite{Buker02,Tikhonov02a,Tikhonov02b,Kohler04,Galperin05,Viljas07b}
remains an important goal for future research.


\acknowledgments

This work was financially supported 
by the Helmholtz Gemeinschaft (Contract No.\ VH-NG-029), 
by the DFG within the Center for Functional Nanostructures, 
and by the EU network BIMORE (Grant No. MRTN-CT-2006-035859). 
F. Pauly acknowledges the funding of a Young Investigator Group at KIT.

\appendix


\section{Simplified formula for the time-averaged current} \label{s.simplecurr}

Consider the expression Eq.\ (\ref{e.photocurr}) for the time-averaged
(or dc) current.  The coefficient $\tau_{RL}^{(k)}(E)$, for example,
is the sum of the transmission probabilities of all transport channels
taking the electron from energy $E$ on the left to energy
$E+k\hbar\omega$ on the right. That is, for $k>0$ ($k<0$) it describes
electron transmission under the absorption (emission) of $k$
photons. Assuming the wide-band approximation and the voltage profile
A, Eq.\ (\ref{e.photocurr}) can be written in the more transparent
forms of Eqs.\ (\ref{e.simplec}) and (\ref{e.simplec2}).  This can be
demonstrated rigorously using the equations of App.\ \ref{s.lightcurr},
but it is instructive to consider the following simpler
derivation. The idea is the same as in the ``independent channel
approximation'' of Ref.\ \onlinecite{Tikhonov02a}.

For now, we allow the ac voltage drops at the $L$ and $R$
lead-molecule interfaces to be asymmetrical. Thus we define the
quantities $\alpha_L$ and $\alpha_R$, satisfying
$\alpha=\alpha_L-\alpha_R$.  Since for profile A there is no voltage
drop on the molecule, electronic transitions only occur at the
lead-molecule interfaces.  Thus the transmission coefficients
$\tau_{RL}^{(k)}(E)$ are given by
\begin{equation}
\tau_{RL}^{(k)}(E)=\sum_{l=-\infty}^{\infty}
\left[J_{l-k}\left(\alpha_R\right)\right]^2
\tau(E+l\hbar\omega)
\left[J_{l}\left(\alpha_L\right)\right]^2,
\end{equation}
where $[J_{l}(\alpha_L)]^2$ is the probability for absorbing
(emitting) $l$ photons on the left interface and
$[J_{l-k}(\alpha_R)]^2$ the probability for emitting (absorbing) $l-k$
photons on the right interface.  The propagation between the
interfaces occurs elastically at the intermediate energy
$E+l\hbar\omega$, according to the transmission function $\tau(E)$.  A
similar expression holds for $\tau_{LR}^{(k)}(E)$.  Using these and
the sum formula $\sum_{k=-\infty}^{\infty}[J_{k}(x)]^2=1$, Eq.\
(\ref{e.photocurr}) leads to
\begin{equation} \label{e.simplec4}
\begin{split}
I(V;\alpha,\omega)
&=\frac{2e}{h}
\sum_{l=-\infty}^\infty \int\upd E \tau(E+l\hbar\omega)\\
&\times 
\left\{
\left[J_{l}\left(\alpha_L\right)\right]^2f_L(E)
-
\left[J_{l}\left(\alpha_R\right)\right]^2f_R(E)
\right\}.
\end{split}
\end{equation}
Equation (\ref{e.simplec}) follows by setting $\alpha_L=\alpha/2$ and
$\alpha_R=-\alpha/2$, and the equivalent form of Eq.\
(\ref{e.simplec2}) follows by changing summation indices and
integration variables.  Similarly, other suggestive forms may be
derived.\cite{Kohler04,Platero04,Pedersen98}
For $x\ll 1$ and $l>0$ one may expand 
$J_{\pm l}(x)\approx (\pm x/2)^l/l! - (\pm x/2)^{l+2}/(l+1)!$. 
This can be used in the limit $\alpha\ll 1$,
$\hbar\omega/\ttt\ll 1$ discussed in the text.


\section{Green's-function method for the time-averaged current} \label{s.lightcurr}

Here we outline the Green-function
method\cite{Datta92,Tikhonov02a,Viljas07a} used for obtaining the
results for voltage profile B.  Consider again the dc current of Eq.\
(\ref{e.photocurr}).  In the case of a harmonic driving field, it is
reasonable to assume the existence of time-reversal invariance, in
which case we have the symmetry\cite{Kohler04}
\begin{equation} \label{e.tr}
\tau_{LR}^{(k)}(E)=\tau_{RL}^{(-k)}(E+k\hbar\omega).
\end{equation} 
The current expression of Eq.\ (8) in Ref.\ \onlinecite{Viljas07a} was
derived under this assumption, and that result can be brought into the
form of Eq.\ (\ref{e.photocurr}).  Using the notation of that
reference,\cite{Note5} the coefficients can be written
\begin{equation} \label{e.coeffs}
\begin{split}
\tau_{RL}^{(k)}(E)&={\Tr}_{\omega}[
\hat{G}(E)\hat{\ga}_{R}^{(k)}(E)\hat{G}^\dagger(E)\hat{\ga}_{L}^{(0)}(E)]
\\
\tau_{LR}^{(k)}(E)&={\Tr}_{\omega}[
\hat{G}(E)\hat{\ga}_{L}^{(k)}(E)\hat{G}^\dagger(E)\hat{\ga}_{R}^{(0)}(E)],
\end{split}
\end{equation} 
where the hats denote the extended ``harmonic'' matrices\cite{Shirley65} and
${\Tr}_\omega$ a trace over them. In particular, $\hat{G}$
is the matrix for the retarded propagator
\begin{equation}
\hat{G}(E)=[(\hat{E}-\mat{H}\hat{1}) -\hat{W} 
-\hat{\se}_{L}(E) -\hat{\se}_R(E) ]^{-1},
\end{equation}
where $\mat{H}$ is the Hamiltonian of the wire in the absence of
voltage profiles. The matrix $\hat{E}$ is defined by
$[\hat{E}]_{m,n}=(E+m\hbar\omega)\delta_{m,n}\mat{1}$, where $m$ and
$n$ are the harmonic indices. Using the wide-band approximation for the
electrodes, the matrices $\hat{\se}_X$ and $\hat{\ga}_X^{(l)}$
are given by
\begin{equation}
\begin{split}
[\hat{\se}_X]_{m,n}(E) &
=\delta_{m,n}\mat{\se}_{X} \\
[\hat{\ga}_X^{(l)}]_{m,n}(E) &
=J_{m-l}(\alpha_X)J_{n-l}(\alpha_X)\mat{\ga}_{X}, \\
\end{split}
\end{equation}
with $X=L,R$ and $\alpha_{L,R}=\pm\alpha/2$.  Here $\mat{\se}_X$ is
the self-energy matrix of lead $X$ (extended to the size of
$\mat{H}$), and $\mat{\ga}_X=-2\im\mat{\se}_X$.  
The matrix
$\hat{W}$ includes the effect of the profiles for the voltage
$V(t)=V+V_{ac}\cos(\omega t)$.  If
$\mat{W}(t)=\mat{W}_{dc}+\mat{W}_{ac}\cos(\omega t)$ is a diagonal
matrix consisting of the onsite energies $\epsilon_p^{(\alpha)}(t)$,
then
\begin{equation}
[\hat{W}]_{m,n}=\mat{W}_{dc}\delta_{m,n}
+\frac{1}{2}\mat{W}_{ac}(\delta_{m-1,n}+\delta_{m+1,n}).
\end{equation}
In this formalism, the time-reversal invariance amounts to
$\hat{G}$ and $\hat{\ga}_{L,R}^{(k)}$ being symmetric, i.e.
$\hat{A}^T=\hat{A}$.
Equation (\ref{e.tr}) can then be proved by using the relations
$[\hat{G}]_{m+k,n+k}(E)=[\hat{G}]_{m,n}(E+k\hbar\omega)$ and
$[\hat{\ga}^{(l)}_X]_{m+k,n+k}(E)=[\hat{\ga}^{(l-k)}_X]_{m,n}(E+k\hbar\omega)$.
We note that $\hat{\ga}^{(l)}_X$ is defined with a different sign 
of $l$ than in Ref.\ \onlinecite{Viljas07a}.



\end{document}